\documentclass[%
twocolumn,
showpacs,
amsmath,amssymb,
aps,
prl,
]{revtex4-1}

\usepackage{subeqnarray}
\usepackage{amsmath}
\usepackage[disable]{todonotes}
\usepackage{graphicx}
\usepackage{natbib}
\usepackage{wasysym}
\usepackage{amssymb}
\usepackage{xcolor}
\usepackage{amsthm}
\usepackage{moreverb} 
\usepackage{textcomp} 
\usepackage{epsfig}
\usepackage{color} 
\usepackage{upgreek}
\usepackage{bbm}

\usepackage{graphics}
\usepackage{dcolumn}
\usepackage{bm}
\usepackage{amsfonts}
\usepackage{textcomp}

\begin{document}

\preprint{APS/123-QED}

\title{An Isotropic Porous Medium Approach to Drag Mitigation Disappearance in Super-Hydrophobic Falling Spheres}

\author{Marco Castagna}
\author{Nicolas Mazellier}%
\email{nicolas.mazellier@univ-orleans.fr}
\author{Pierre-Yves Passaggia}
\author{Azeddine Kourta}
\affiliation{%
University of Orl\'eans, INSA-CVL, PRISME, EA 4229, 45072, Orl\'eans, France 
}%




\date{\today}

\begin{abstract}
In this Letter, the falling of super-hydrophobic spheres is investigated experimentally at low Reynolds numbers. In particular, we show that super-hydrophobic coatings become ineffective at reducing drag unlike predicted by theoretical and numerical approaches. A time scale analysis describing both Marangoni-induced stresses and air/liquid interface deformation shows that these mechanisms are unlikely to account for the slippage effect disappearance observed in our study. Instead, we propose a simple model based on an isotropic porous medium approach, derived to account for losses induced by the motion of the gas encapsulated around the sphere. The key parameter of this mechanism is found to be the surface tortuosity, whose range estimated from microscopic surface imaging corroborate those predicted by our scaling analysis and previous studies.
\end{abstract}

\maketitle

More than a century after the seminal work of Stokes \cite{Stokes1851}, bluff-body drag and in particular its control remains an open question. Combined efforts from both theoretical fluid dynamics \cite{Albano1975} and material sciences \cite{Quere2008} demonstrated the potential of wall slip type surfaces towards drag reduction. Ideally, air entrapment between the liquid and the solid induces a local slip condition which may favourably decrease drag by momentum transfer at the air/liquid interface \cite{McHale2011}. This can be practically achieved in a passive manner combining surface texturing and chemical repellence, leading to the so-called super-hydrophobic (SH) surfaces \cite{Rothstein2010}. Despite the potential of this emerging technology, the discordant results between theory/simulations \cite{Willmott2009,Gruncell2013} and the few available experimental studies \cite{Byon2010,Ahmmed2016,Castagna2018} on SH bluff-bodies suggest that physical modelling of SH surfaces still has to be improved. In the following we focus on the low Reynolds number range where mild bluff-body-drag reduction was experimentally reported \cite{Byon2010,Ahmmed2016}.

In recent years, research groups tried to explain this limitation of SH drag reduction ability by taking into account different effects. Peaudecerf et al. \cite{Peaudecerf2017} investigated the Marangoni flow generated as a result of the surface tension gradient due to the build-up of contaminants on the air/liquid interface. They showed that surfactant-induced stresses can become significant, even at very low contaminants concentrations, potentially yielding a no-slip boundary condition over the flat air/liquid interface. These results were extended by Song et al. \cite{Song2018} who demonstrated that the Marangoni effect strength is dependent on the roughness arrangement. Comparing closed cavities and continuous grooves, they showed that the prevention of the surfactant build-up in the later case induces a negligible surface tension gradient, restricting thereby the adverse effect of the Marangoni flow. In addition, they observed that the flow over closed cavities is sensitive to the shape of the air/liquid interface. Convex air/liquid interfaces were found to maximise the slip velocity with respect to the concave counterpart. Following the work of \cite{McHale2011}, Gruncell et al. \cite{Gruncell2013} analysed the influence of roughness elements of the momentum transfer occurring at the air/liquid interface. They showed that solid fractions resembling the SH surfaces used in this study yield detrimental effects to a potential drag increase. It is worth noting that their study only considers an axisymmetric bluff-body in the approximation of non-deformable air/liquid interfaces.

\begin{figure}[ht!]
\hspace{-0mm}\scalebox{0.28}{\Large\input{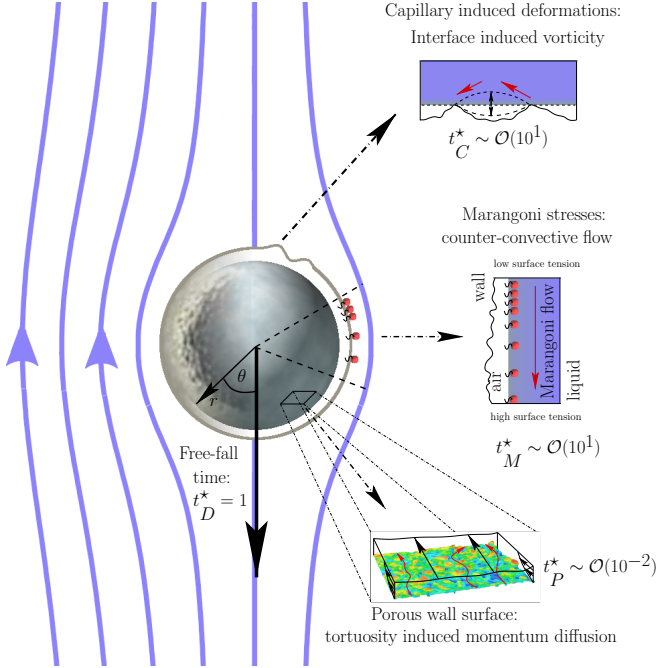}}%
\vspace{-4mm}
\caption{Schematics summarizing to the right, the mechanisms influencing the performances of super-hydrophobicity in affecting drag, where each mechanism is characterized by its characteristic time $t^{\star}$, normalized by the free-fall time $t_D$.}
\label{schematic}
\end{figure}

We therein consider a new effect that was not accounted for in these previous studies. Using an isotropic porous media approximation, we model the resistance induced by the air motion around the roughness elements. To this end, we designed an experiment where Marangoni-induced stresses and air/liquid interface deformation are unlikely to be the predominant mechanisms. Each of the aforementioned mechanisms is depicted in Fig. \ref{schematic}, which summarizes the state of the art in SH modelling at low Reynolds numbers.

Falling sphere experiments were performed in a transparent tank ($100\times100$  $mm^{2}$ square cross-section and 650 mm height) filled with glycerine (see \cite{Castagna2018} and Supplementary Material for details). 
%
The glycerine density $\rho_{l}$ and dynamic viscosity $\mu_{l}$ were evaluated using the empirical formula proposed by \cite{Cheng2008} and $\mu_{l}$ was further experimentally verified with a Fungilab{\texttrademark} viscometer. Stainless steel spheres with nominal diameters $d$ equal to 5, 8 and 10 mm were taken as reference.
The trajectory of the falling sphere was recorded by a Phantom V341 high-speed camera at a 2560$\times$1100 px$^{2}$ resolution, resulting into a conversion factor of 0.3 mm px$^{-1}$.
The recording frame rate was adapted to the sphere falling velocity, ranging from 200 fps for the $d=5$ mm up to 800 fps for the $d=10$ mm spheres.
%
%
The 
terminal velocities $V_{\infty}$ were evaluated in the range $0.07-0.23$ ms$^{-1}$, which correspond to a terminal Reynolds number ($Re_{\infty}=\rho_{l}V_{\infty}d/\mu_{l}$) lying within $0.3-2.5$. 
Confinement effects induced by the finite size of the test section were taken into account
following the technique proposed by Di Felice \cite{DiFelice1996} (see Supplementary Material for full details). Moreover, the SH spheres were produced by a spray method technique suitable for macroscopic applications, 
and is
fully described in \cite{Castagna2018}. The three produced SH coatings will be indicated hereinafter SH-1, SH-2 and SH-3 in order of increasing root-mean square surface roughness $\lambda$ values. The employed manufacturing procedure resulted into a randomly distributed surface roughness. Details of the manufactured coatings are available in the Supplementary Material.

\begin{figure}
\centerline{\includegraphics[width = 1 \columnwidth]{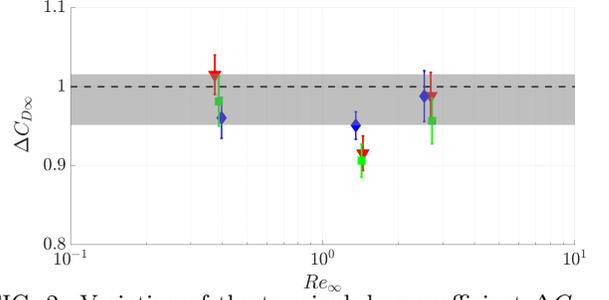}}
\vspace{-5mm}
\caption{Variation of the terminal drag coefficient $\Delta C_{D\infty}$ of the SH spheres with respect to the corresponding reference sphere, as a function of the terminal Reynolds number $Re_{\infty}$. $\textcolor{red}{\blacktriangledown}$, SH-1 coating. $\textcolor{blue}{\blacklozenge}$, SH-2 coating. $\textcolor{green}{\blacksquare}$, SH-3 coating. The error bars represent the 95$\%$ confidence level. The dashed line outlines the unitary reference value. The grey-shaded rectangle indicates the range of the drag variation predicted by the porous medium model.}
\label{fig:2}
\end{figure}

The glycerine-filled tank was tall enough to allow 
every
sphere to reach 
its
respective terminal velocity $V_{\infty}$, 
with negligible transverse motions.
The terminal drag coefficient is therefore computed by the following relation:
\begin{equation}
C_{D\infty}= 4dg\left(\zeta-1\right)/\left(3V_{\infty}^{2}\right),
\end{equation}
where $g$ is the acceleration due to gravity and $\zeta=\rho_{s}/\rho_{l}$ is the density ratio between sphere and glycerine, 
in the range $4.8-6.1$. Fig. \ref{fig:2} shows the variation of the terminal drag coefficient $\Delta C_{D\infty} = C_{D\infty}^{SH}/C_{D\infty}^{ref}$ as a function of $Re_{\infty}$, highlighting the absence of a noteworthy drag reduction due to the SH coatings. The experimental data show a possible drag reducing effect of SH coatings limited to $<10\%$. This value is considerably lower than predictions by analytical studies and numerical simulations \cite{McHale2011,Gruncell2013}. Moreover, no considerable effect of the surface roughness can be derived, since the error bars (95$\%$ confidence level) of the different coatings tend to overlap.

We summarize the mechanisms shown in Fig. \ref{schematic} according to their characteristic time and show that they become irrelevant with respect to the free-fall time scale.
A scaling analysis of the characteristic Marangoni time $t_{M}$ and drop time $t_{D}$ 
shows that the time scale of the sphere fall is so short that the build up of eventual surfactants cannot be
a inhibiting mechanism for SH drag mitigation.
In fact, equating drag and gravity/buoyancy forces at steady state, we can derive an expression of the terminal velocity which writes:
\begin{equation}
V_{\infty}=\frac{g\left(\zeta -1 \right)\rho _{l}d^{2}}{18\mu _{l}}=\frac{V_{D}}{18}\frac{\rho _{l}V_{D}d}{\mu _{l}}=\frac{V_{D}Re_{D}}{18},
\end{equation}
where $V_{D}=\sqrt{\left(\zeta -1 \right)gd}$ is chosen as the scaling velocity in order to account for gravity/buoyancy effects \cite{Jenny2004}, and the Reynolds number $Re_{D}$ is defined accordingly. 
Similarly, the scaling time of the drop is defined as $t_{D}=\sqrt{d/\left( \left(\zeta-1\right)g\right)}$ and can be compared to the Marangoni time $t_{M}\sim d / u_{sl}$, where the diameter $d$ and the slip velocity $u_{sl}$ are chosen as the characteristic length and velocity of the air layer. In fact, the SH coatings investigated in our work likely result into a continuous air layer rather than a closed groove geometry.
%
We model the slip velocity as follows:
\begin{equation}
{u}_{sl} \sim \lambda \tau / \mu_{l},
\label{eq:Navcond}
\end{equation}
where the stress $\tau$ at the wall is linked to the slip velocity $u_{sl}$ by the slip length, which as a first approximation is taken equal to the rms surface roughness $\lambda$ \cite{Zampogna2018}.
We can define the total SH sphere surface as:
\begin{equation}
S_{SH} = 4\pi \left(r+\lambda \right)^{2} \approx 4\pi r^{2} \left(1+2\lambda/r  \right),
\label{eq:SSH}
\end{equation}
where $r=d/2$ is the nominal radius of the reference sphere, and the expression was simplified 
given
$d \gg \lambda$. 
The wall stress $\tau$ can thus be expressed as:
\begin{eqnarray}
\tau \sim \frac{F_{SH}}{S_{SH}} \approx \frac{6\pi \mu_{l}\left(r-\lambda \right)V_{\infty}}{4\pi r^{2}\left(1+\frac{2\lambda}{r} \right)} \approx \nonumber \\
\approx \frac{3\mu_{l}V_{\infty}}{2r^{2}}\left(r-\lambda \right)\left(1-\frac{2\lambda}{r}\right) \approx \frac{3}{2}\mu_{l}V_{\infty}\frac{r-3\lambda}{r^{2}},
\label{eq:tau}
\end{eqnarray}
where $F_{SH}$ is the total Stokes' force acting on the SH sphere.
Substituting Eq. \ref{eq:tau} into Eq. \ref{eq:Navcond} we obtain:
%
%
\begin{equation}
u_{sl} \sim \frac{\lambda \tau}{\mu_{l}} \approx \frac{\lambda}{\mu_{l}}\frac{3}{2}\mu_{l}V_{\infty}\frac{r-3\lambda}{r^{2}} \approx \frac{3}{2}\frac{\lambda}{r}V_{\infty}.
\label{eq:uslfinal}
\end{equation}
The $u_{sl}$ expression 
allows for deriving
a scaling for
the strength of the characteristic time of Marangoni stresses with respect to the drop time (the superscript $^{\star}$ indicates normalisation with respect to $t_{D}$):
\begin{equation}
t_{M}^{\star} \sim \frac{d \sqrt{\left(\zeta -1\right)g}}{u_{sl}\sqrt{d}} \approx \frac{6}{Re_{D}} \frac{d}{\lambda} \approx \mathcal{O}\left(10^{1}\right).
\end{equation}
The Marangoni time $t_{M}$ is thus at least one order of magnitude larger than the characteristic drop time $t_{D}$. 

The negligible drag reducing ability of SH spheres may be linked to the deformation of the air-liquid interface, which was extensively investigated in \cite{Castagna2018,Song2018}.
Similarly to the Marangoni flow case, we can define a capillary time $t_{C}$ by relating the sphere length scale $d$ to a characteristic velocity $V_{C} \sim \sqrt{\gamma /\left(\rho_{l}d \right)}$, with $\gamma$ the air/liquid surface tension. The ratio between the capillary and the drop times 
gives:
\begin{equation}
t_{C}^{\star} \sim \frac{d}{V_{C}t_{D}} \approx \sqrt{\rho_{l} dV_{D}^{2}/\gamma} \approx \sqrt{Re_{D}Ca_{D}} \approx \mathcal{O}\left(10^{1}\right),
\end{equation}
where the capillary number is defined as $Ca_{D}=\mu_{l}V_{D}/\gamma \approx \mathcal{O}\left(10^{1}\right)$. The capillary time $t_{C}$ 
also results in approximately one order of magnitude larger than the characteristic drop time $t_{D}$, meaning that an interface deformation is unlikely to occur. This fact was further verified by high-magnification tests whose objective was to detect a possible deformation of the air layer during the drop. A Sigma 180 mm Macro objective was used, which let us reach a 0.05 mm px$^{-1}$ resolution. No macroscopic deformation of the air layer was noticed during the drop of all the investigated spheres: the movement and deflection, if present, are so limited that are not detectable with the current experimental set-up.

An additional
mechanism can 
be related to the ability of the external flow to transfer momentum across 
the air/liquid interface. This diffusion-based mechanism
acts
inside the air layer and can be described by 
the characteristic time, $t_{P} \sim \lambda / \nu_{a}$, 
which connects the height of the air layer $\lambda$ with the air kinematic viscosity $\nu_{a}$. The comparison with the drop time thus yields:
\begin{equation}
t_{P}^{\star} \sim \frac{\lambda^{2}V_{D}}{\nu_{a}d} \approx \left(\frac{\lambda}{d} \right)^{2} Re_{D}\frac{\nu_{l}}{\nu_{a}} \approx \mathcal{O}\left(10^{-2}\right),
\end{equation}
where $\nu_{l}=\mu_{l}/\rho_{l}$ is the liquid kinematic viscosity. 
This laminar diffusion mechanism erodes momentum between the air layer and the rough surface of the sphere, where the analogy with a flow across a porous media can be made.

We therefore adopt this framework
to explain the absence of a significant SH drag reducing effect in the present laminar regime. The basic idea is to show that if the lubricating effect due to presence of the air layer ($\delta F_{sl}$) can be partially/completely balanced by the resistance induced by the surface roughness ($\delta F_{por}$), this may lead to a limited or negligible drag decrease in the case of SH spheres. 
The total force acting on the sphere in such Stokes' flow could be modelled as:
\begin{equation}
F_{SH} = F_{St}-\delta F_{sl}+\delta F_{por},
\label{eq:FSHnew}
\end{equation}
where $F_{St}=6\pi \mu_{l} r V_{\infty}$ is the expression of the Stokes' drag force valid for a no-slip sphere. We 
initially follow the approach of Willmott \cite{Willmott2009} who performed analytical calculations of ``Janus" spheres (partially SH spheres) in laminar flow, evaluating the lubricating force as a function of the surface coverage of the SH coating. Considering a fully SH sphere, the first order term of the expansion of the lubricating force can be written as \cite{Willmott2009}:
\begin{equation}
\delta F_{sl} = 6\pi\mu_{l}\lambda V_{\infty}.
\label{eq:lubr}
\end{equation}
An expression for the $\delta F_{por}$ contribution can be retrieved, considering an elementary volume $\mathcal{V}$ of characteristic length $L$ and height $\lambda$ inside the air layer on the rough surface. The elementary pressure loss across $\mathcal{V}$ can be expressed, following a Darcy (or Fanning) formalism, through the friction factor $f$ as follows:
\begin{equation}
\Delta p = fL\lambda^{-1}\rho_{a}u_{sl}^{2}/2.
\label{eq:deltapf}
\end{equation}
Introducing the permeability $\kappa$ of the porous medium, one can show that:
\begin{equation}
f = 2 \lambda^{2}\kappa^{-1}Re_{P}^{-1},
\label{eq:fk}
\end{equation}
where $Re_{P}=\rho_{a}u_{sl}\lambda/\mu_{a}$ is the characteristic Reynolds number inside the air layer. 
Under the hypothesis of an isotropic porous medium, the permeability can be explicitly expressed as follows \cite{Bear1988}: 
\begin{equation}
\kappa = \phi \lambda ^{2}/\left(96 T^{2}\right),
\label{eq:kappa}
\end{equation}
where $\phi$ is the porosity of the medium, which can be evaluated as the ratio between the void volume and the total volume of the medium, and $T$ is the tortuosity, defined as the ratio between the actual length covered by the air flow and the total porous medium length. The drag over the elementary volume can be obtained by multiplying the pressure loss in eq. (\ref{eq:deltapf}) by the elementary cross section such that:
\begin{equation}
D = \Delta p L \lambda = f L^{2}\rho_{a}u_{sl}^{2}/2,
\label{eq:D}
\end{equation}
which can be further divided by the elementary surface $L^{2}$ to 
provide a measure of 
a drag per unit surface area. The latter can be scaled over the whole sphere surface and reads:
\begin{equation}
\delta F_{por} = f\rho_{a}u_{sl}^{2}4\pi r^{2}/2 = 384T^2\phi^{-1}\pi \mu_{a}u_{sl}r^{2}\lambda^{-1}.
\label{eq:deltaFporl}
\end{equation} 
The ratio between the beneficial slip effect and the detrimental roughness contribution  
finally writes:
\begin{equation}
\frac{\delta F_{por}}{\delta F_{sl}} = \frac{384T^{2}\pi \mu_{a}u_{sl}r^{2}}{6\pi \mu_{l}\lambda V_{\infty} \phi \lambda} \approx \frac{96T^2}{\phi}\left(\frac{\mu_{a}}{\mu_{l}}\right)\left(\frac{r}{\lambda}\right).
\label{eq:deltaFratio}
\end{equation}
In order for the quantity in eq. \ref{eq:deltaFratio} to be $\approx \mathcal{O}\left(10^{0}\right)$, the following expression must be valid:
\begin{equation}
T \approx \sqrt{\frac{\phi}{96}\left(\frac{\mu_{l}}{\mu_{a}}\right)\left(\frac{\lambda}{r}\right)} \approx \mathcal{O}\left(10^{0} - 10^{1}\right). 
\label{eq:T}
\end{equation}
This range corresponds to the typical empirical values reported in literature (\cite{Bear1988}, and references therein) and was experimentally verified with the aid of 3D confocal microscopy images (see Supplementary Material). Fig. \ref{fig:3} shows a binary representation of one of the analysed SH coatings, where an A$^{*}$ path detection algorithm was implemented \cite{Ueland2017}. The value $T=1.6$ evaluated in Fig. \ref{fig:3}(a) in the optimal path case can quickly be raised to higher values ($T=3.2$ in Fig. \ref{fig:3}(b)) by selecting non-optimal paths. The grey-shaded area in Fig. \ref{fig:2} testifies the quite good agreement between the experimental data and the model, the latter being able to predict both slight drag augmentation and reduction.


\begin{figure}
\centerline{\includegraphics[width = 1 \columnwidth]{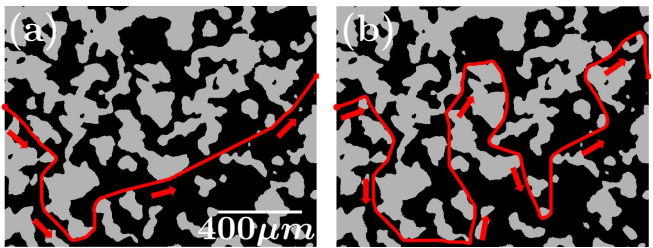}
\put(-244,78){\textbf{\textcolor{white}{\textbf{\Large(a)}}}}\put(-118,78){\textbf{\Large\textcolor{white}{(b)}}}
}
\vspace{-3mm}
\caption{Path detection on a binarised 3D confocal image of a SH-2 coating flat plate. (a) optimal path, (b) non-optimal path. The black and gray regions indicate the areas below and above the rms surface roughness $\lambda$, respectively. The red dot and square indicate the starting and target points, respectively. The red arrows indicate the path direction.}
\label{fig:3}
\end{figure}

In conclusion, our SH falling sphere experiments show a negligible effect on drag of several SH coatings at low Reynolds numbers. A time scale analysis suggests that motions of the air encapsulated inside the surface roughness elements are a suitable mechanism to explain the loss of hydrodynamic performances. The resulting drag variation is modelled via a porous medium approach, which found the tortuosity to be a key parameter 
to quantitavely
describe the mechanism. This model predicts that tortuosity should be minimized in order to be effective at reducing the resistance induced by surface roughness. The tortuosity range estimated by the model agrees also well with the experimental data and the available literature.
Further work should be performed to deepen the knowledge on the role of the tortuosity parameter on SH coatings hydrodynamic performance. A compromise between 
scalable industrial spray coating
and expensive small scale regular geometry techniques should be found in the attempt to reproduce on large industrial applications the outstanding performance of the lotus leaves (see e.g. \cite{Ensikat2011}). Finally, it must be noted that some of the conclusions proposed above could still be valid at higher Reynolds numbers, since works available in literature (see e.g. \cite{Daniello2009}) have shown the key role of the relative viscous sub-layer thickness with respect to the SH coating features size in turbulent flows. However, the evolution of the ratio between the detrimental surface roughness resistance and the beneficial slip lubricating effect with increasing Reynolds number is not predictable by the present model and deserves further analysis.     


This work was supported by the Direction G\'en\'erale de l'Armement (DGA), Minist\`ere de la D\'efense, R\'epublique Fran\c caise and the Agence Nationale de la Recherche (ANR) through the Investissements d'Avenir Program under the Labex CAPRYSSES Project (ANR-11-LABX-0006-01).

\vspace{-8mm}
\bibliography{biblio}

\end{document}


\preprint{APS/123-QED}

\title{Supplemental Material}

\author{Marco Castagna}
\author{Nicolas Mazellier}%
\email{nicolas.mazellier@univ-orleans.fr}
\author{Pierre-Yves Passaggia}
\author{Azeddine Kourta}
\affiliation{%
University of Orl\'eans, INSA-CVL, PRISME, EA 4229, 45072, Orl\'eans, France 
}%




\date{\today}
             
\maketitle

\section{Experimental Set-Up}
\begin{figure}
\centerline{\includegraphics[width = 0.9 \columnwidth]{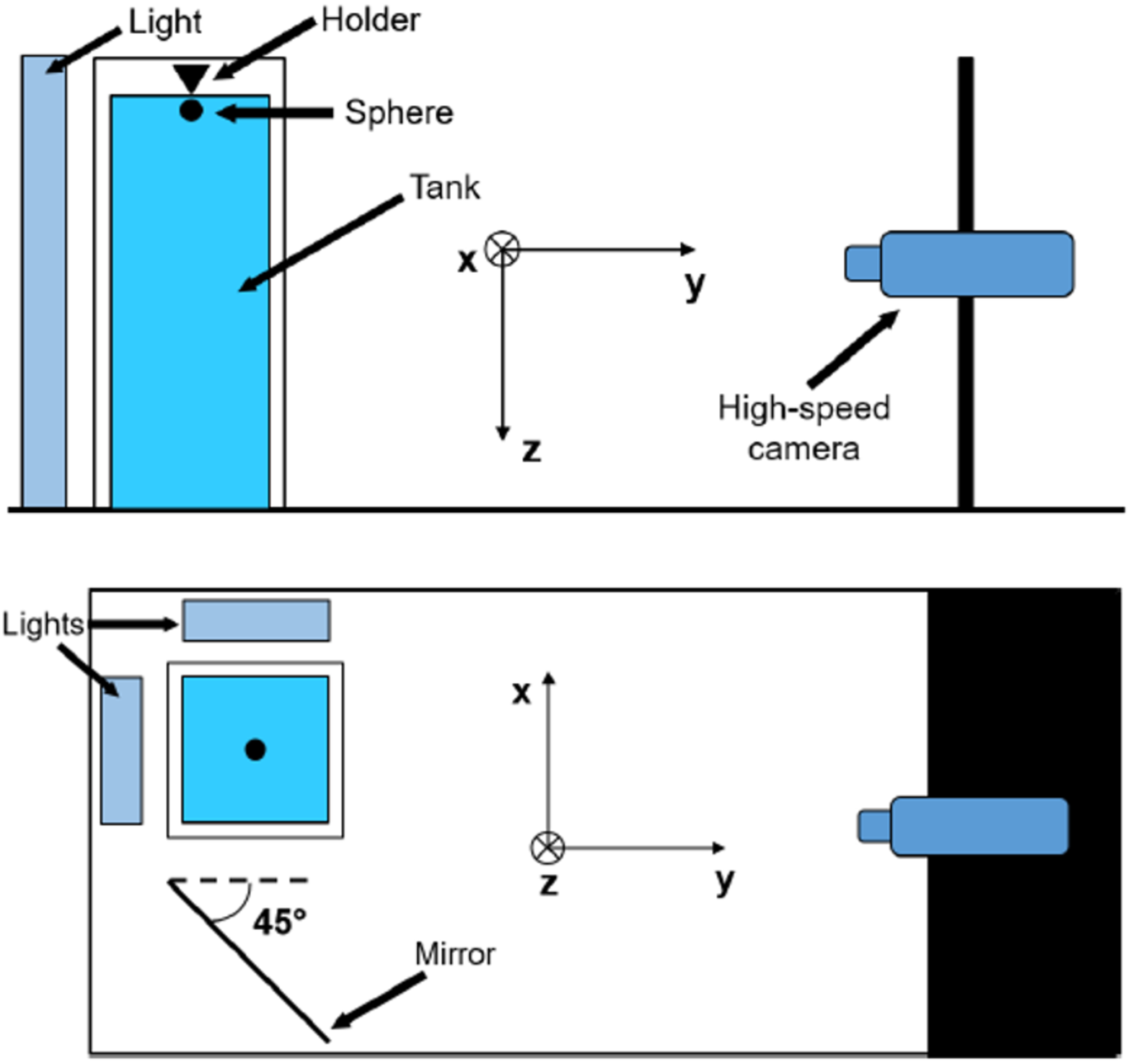}
\put(-235,110){\textbf{\textcolor{black}{\textbf{\normalsize(a)}}}}\put(-235,5){\textbf{\normalsize\textcolor{black}{(b)}}}
}
\caption{The falling sphere experimental set-up: (a) side view, (b) top view.}
\label{fig:1Suppl}
\end{figure}

The falling sphere experiments were performed in a transparent tank ($100\times100$  $mm^{2}$ square cross-section and 650 mm height) filled with glycerine (see Fig. \ref{fig:1Suppl}). Full details were provided in \cite{Castagna2018} and are summarised in the following. The results presented in this study are the average of five independent trials for each sphere. A settling time of at least one hour was imposed between two tests in order to assure the fluid was completely quiescent. Longer settling times (up to one night) were tested, showing no appreciable differences on the post-processed results. All the tests were performed at a room temperature equal to $20\pm1$ $^{\circ}C$, which was verified by a thermocouple dipped in the glycerine. At the beginning of the test, an electromagnetic holder was used to gently dip the spheres below the glycerine surface in order to assure a null velocity at release time. Different recording frame rates were used since the falling velocity strongly varied with the sphere diameter. The largest frame rate of 800 fps was used for the $d=10$ mm spheres, while the falling tests of the $d=5$ mm spheres were recorded at 200 fps. The presence of a mirror placed at 45$^{\circ}$ with respect to the tank enabled the simultaneous recording of the front and side views with a single camera, thus providing information on the three-dimensional (3D) displacement of the falling sphere. The post-processing of the recorded videos which enabled the reconstruction of the 3D displacement of the falling sphere was performed with the commercial software MATLAB\textsuperscript{\textregistered}. A cross-correlation code was developed to achieve sub-pixel accuracy by means of a Gaussian fit of the correlation peak \citep{Raffel2007}. The accuracy was estimated to be approximately $\pm$ $0.06$ px, that is lower than $0.4\%$ of the smallest investigated diameter.

Confinement effects induced by the finite size of the test section were taken into account applying the following relation proposed by Di Felice \cite{DiFelice1996}:
\begin{equation}
\frac{V_{\infty}^{bou}}{V_{\infty}^{unb}}=\left(\frac{1-\delta}{1-0.33\delta}\right)^{\alpha},
\label{eq:Vbound}
\end{equation}
where the ratio of the bounded terminal velocity $V_{\infty}^{bou}$ with respect to the unbounded $V_{\infty}^{unb}$ counterpart depends on the blockage factor $\delta=d/D_{tank}^{eq}$. In the latter expression $D_{tank}^{eq}=2l/\sqrt{\pi}$ ($l$ being the tank width) is the equivalent diameter of the tank. In this study, the blockage factor $\delta$ is lower than 0.09. Finally, the exponent $\alpha$ in Eq. \ref{eq:Vbound} was evaluated by the following expression \cite{DiFelice1996}:
\begin{equation}
\frac{3.3-\alpha}{\alpha-0.85}=0.27Re_{\infty}^{0.64}.
\end{equation}

\section{SH coatings}

\begin{figure}
\centerline{\includegraphics[width = 0.8 \columnwidth]{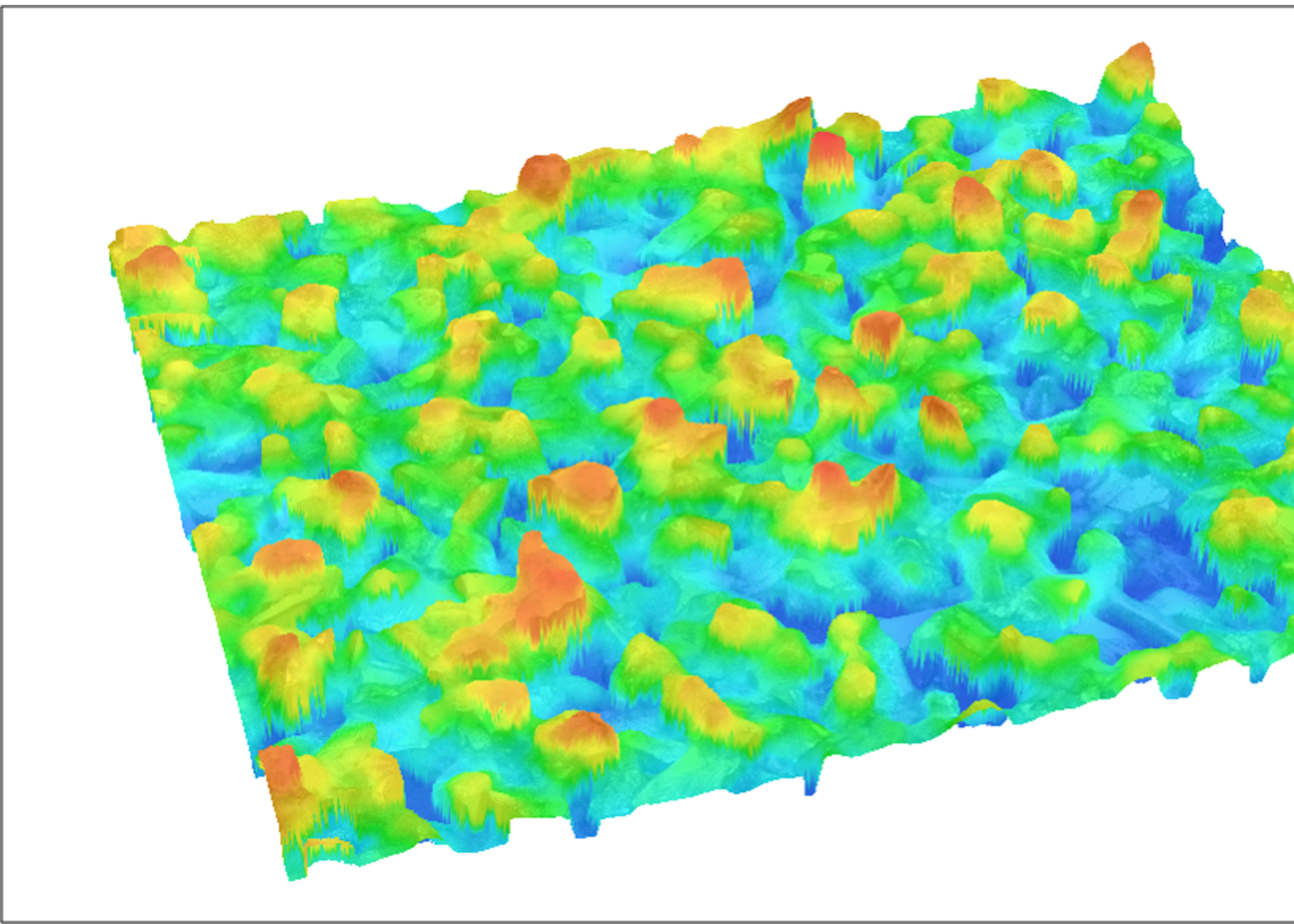}}
\caption{Confocal microscopy analysis of a portion ($1.5\times1.0$ mm) of a flat plate covered by the SH-2 coating. The color scale (blue-to-red) indicates the surface roughness (0 - 170 $\mu$m). Notice the completely random spatial distribution of the powder particles.}
\label{fig:2Suppl}
\end{figure}

The SH spheres were produced by applying a SH coating over the reference spheres via a spray method, using a commercially available SH painting (Ultra-Ever Dry\textsuperscript{\textregistered}). The manufacturing procedure is fully detailed in \cite{Castagna2018}. The deposition of carbon-based powders with different nominal particle sizes enabled us to control the surface texture. The three produced SH coatings will be indicated hereinafter SH-1, SH-2 and SH-3 in order of increasing root-mean square surface roughness $\lambda$ values. The latter was evaluated by 3D confocal microscopy measurements over SH flat plates (see Fig. \ref{fig:2Suppl}), with values in the range $25-142$ $\mu$m (see Tab. \ref{tab:1Suppl}). Contact angle measurements were performed with a digital goniometer via the sessile drop technique. Glycerine drops with a volume of 6 $\mu L$ were deposited over SH horizontal flat plates providing static contact angles ($\theta_{s}$) in the range $137-156$ degrees. The highest $\theta_{s}$ was obtained with the SH-1 coating, while a $\lambda$ increase determined a $\theta_{s}$ reduction, in good agreement with the findings reported in \cite{Nilsson2010} for surface roughness in the same range of our study. The same approach was followed to evaluate the roll-off ($\theta_{r}$) and the hysteresis ($\theta_{h}$) angles by tilting the SH flat plates at a controlled rate of 0.5 deg/s. The $\theta_{r}$ angle, which was considered in this study as the lowest tilt angle that caused the drop to roll-off, reached the highest value of $5.4$ deg. in the SH-3 case, whereas a $\lambda$ decrease determined a $\theta_{r}$ reduction down to $2.6$ deg. for the SH-1 coating. The $\theta_{h}$ angle, defined as the difference between the advancing and receding angles at the time instant where the drop starts to roll-off \cite{Gao2006}, reached the highest value of $11.6$ deg. in the SH-3 case, while it decreased down to $4.9$ deg. for the SH-1 coating with the lowest $\lambda$.
The SH coatings determined a $d$ increase (measurement accuracy 10 $\mu$m) with respect to the reference spheres, with a maximum $+ 10\%$ in the SH-3 case. The masses of the reference spheres (measurement accuracy 0.1 mg) were 0.51, 2.07 and 4.04 g for the spheres with diameter 5, 8 and 10 mm respectively. The SH coatings determined a mass increase up to $+6\%$ in the SH-3 case. Full details of the experimental set-up, SH coatings manufacturing procedure, sphere properties and post-processing techniques are presented in \cite{Castagna2018}. 

\begin{table}
\begin{ruledtabular}
\begin{tabular}{ccccc}
 & SH-1 & SH-2 & SH-3 \\ 
$\lambda$ [$\mu m$] & 25 $\pm$ 4 & 74 $\pm$ 12 & 142 $\pm$ 23 \\ 
$\theta_{s}$ [deg.] & 156.4 $\pm$ 5.8 & 147.3 $\pm$ 7.6 & 136.7 $\pm$ 12.0 \\ 
$\theta_{r}$ [deg.] & 2.6 $\pm$ 0.8 & 2.3 $\pm$ 0.8 & 5.4 $\pm$ 2.6 \\ 
$\theta_{h}$  [deg.] & 4.9 $\pm$ 3.1 & 5.5 $\pm$ 3.9 & 11.6 $\pm$ 4.9 \\ 
\end{tabular}
\end{ruledtabular}
\caption{Properties of the manufactured SH coatings. $\lambda$, root-mean square surface roughness. $\theta_{s}$, static contact angle. $\theta_{r}$, roll-off angle. $\theta_{h}$, hysteresis angle. The reported uncertainties represent the 95$\%$ confidence level.}
\label{tab:1Suppl}
\end{table}

\vspace{-6mm}
\bibliography{biblio}